\documentstyle[12pt]{article}
\advance\textheight by \topskip
\makeatletter

\@addtoreset{equation}{section}
\makeatother

\newcommand{\BE}{\begin{equation}}
\newcommand{\EE}{\end{equation}}
\newcommand{\BA}{\begin{eqnarray}}
\newcommand{\EA}{\end{eqnarray}}

\def\no{\nonumber}
\def\bi{\bibitem}

\def\ap{\alpha ' }
\def\B{\beta}
\def\BH{\beta_{\scriptscriptstyle H}}

\def\th{\vartheta}
\def\T2{|T|^2}
\def\Eb{\overline{E}}
\def\p{\partial}
\def\a{\alpha}
\def\D#1{D#1-$\overline{\textrm{D#1}}$}
\def\NS9{NS9B-$\overline{\textrm{NS9B}}$}
\def\Cm{C_{\scriptscriptstyle -1}}
\def\Dm{D_{\scriptscriptstyle -1}}
\def\Ch{C_{\scriptscriptstyle - \frac{1}{2}}}
\def\Dh{D_{\scriptscriptstyle - \frac{1}{2}}}
\def\v{{\cal V}}

\input epsf

\begin{document}

\rightline{KUNS-1814}
\rightline{hep-th/0212063}

\vspace{.8cm}
\begin{center}
{\large\bf Brane-Antibrane Systems at Finite Temperature

and Phase Transition near the Hagedorn Temperature
}

\vskip .9 cm

{\bf Kenji Hotta,}
\footnote{E-mail address: khotta@gauge.scphys.kyoto-u.ac.jp}

Department of Physics, Kyoto University, Kyoto 606-8502, JAPAN
\vskip 1.5cm

\end{center}
\vskip .6 cm
\centerline{\bf ABSTRACT}
\vspace{-0.7cm}
\begin{quotation}

In order to study the thermodynamic properties of brane-antibrane systems, we compute the finite temperature effective potential of tachyon $T$ in this system on the basis of boundary string field theory. At low temperature, the minimum of the potential shifts towards $T=0$ as the temperature increases. In the {\D{9}} case, the sign of the coefficient of $\T2$ term of the potential changes slightly below the Hagedorn temperature. This means that a phase transition occurs near the Hagedorn temperature. On the other hand, the coefficient is kept negative in the {\D{p}} case with $p \leq 8$, and thus a phase transition does not occur. This leads us to the conclusion that only a {\D{9}} pair and no other (lower dimensional) brane-antibrane pairs are created near the Hagedorn temperature. We also discuss a phase transition in {\NS9} case as a model of the Hagedorn transition of closed strings.

\end{quotation}

\normalsize
\newpage

\section{Introduction}
\label{sec:Intro}

Recently, significant effort has been devoted to studying non BPS configurations of branes such as brane-antibrane pairs and non BPS D-branes (for a review see e.g. \cite{Ohmori}). If we consider a coincident brane-antibrane pair, the spectrum of open strings on it contains a complex tachyon field $T$. This indicates that the brane-antibrane system is unstable, and that the tachyon rolls down from the perturbative vacuum of open strings. If we assume that the tachyon potential has a non-trivial minimum, it is expected that the tachyon falls into it. Sen conjectured that the negative energy density from the tachyon potential exactly cancels the sum of the tension of the brane and the antibrane \cite{Senconjecture}. This means that the brane-antibrane pair disappears and the energy density of this configuration vanishes at the potential minimum. In fact, the tachyon potential on this configuration calculated by using cubic string field theory approves Sen's conjecture \cite{CSFT}.

The tachyon potential on the {\D{p}} system in type II string theory was also computed on the basis of boundary string field theory (BSFT) \cite{BSFT1} \cite{BSFT2} by Minahan and Zwiebach \cite{tachyon1} and by Kutasov, Marino and Moore \cite{tachyon2}, and its qualitative features agree with Sen's conjecture. If we denote the complex scalar tachyon field by $T$, then the potential is given by
\BE
  V(T) = 2 \tau_p \v \exp (-8 \T2),
\EE
where $\v$ is the volume of the system that we are considering, and $\tau_p$ denotes the tension of a Dp-brane, which is defined by
\BE
  \tau_p = \frac{1}{(2 \pi)^p {\ap}^{\scriptscriptstyle \frac{p+1}{2}} g_s},
\label{eq:tension}
\EE
where $g_s$ is the coupling constant of strings. This potential has the minimum at $|T| = \infty$. We adopt the natural unit $c= \hbar =1$ but explicitly keep the slope parameter $\ap$.

It is interesting to investigate the finite temperature system of {\D{p}} pair. This is because there is a possibility that $T=0$ becomes a stable minimum at high temperature, in analogy with high temperature symmetry restoration in field theories. For this purpose, we must compute the finite temperature effective potential of the brane-antibrane system based on BSFT, when the complex tachyon field develops its vacuum expectation value.

Danielsson, G\"{u}ijosa and Kruczenski evaluated the finite temperature effective potential when there is only the tachyon field \cite{LowTtach}. The effective action of tachyon field at zero temperature can be deduced from the tree level BSFT of open superstrings as \cite{tachyon1} \cite{tachyon2}
\BE
  S = -16 \tau_p \int d^{p+1} x \left[ 2 \ap e^{-8 \T2} \p T \p T^*
    + \frac{1}{8} e^{-8 \T2} \right].
\label{eq:taceffaction1}
\EE
In order to obtain the standard form of the kinetic term, we perform a field redefinition
\BA
  \phi = \frac{1}{\sqrt{\pi}} \int_{0}^{2T} e^{-|u|^2} du. \no
\EA
Then the action (\ref{eq:taceffaction1}) becomes
\BE
  S = -8 \pi \ap \tau_p \int d^{p+1} x \left[ \p \phi \p \phi^*
    + \frac{1}{4 \pi \ap} e^{-8 |T( \phi )|^2} \right].
\label{eq:taceffaction2}
\EE
The mass square of the field $\phi$ is given by
\BE
  m^2 = \frac{\p^2}{\p \phi \p \phi^*}
    \left( \frac{1}{4 \pi \ap} e^{-8 |T( \phi )|^2} \right)
      = \frac{1}{\ap} \left( |T( \phi )|^2 - \frac{1}{2} \right),
\label{eq:tachyonmass}
\EE
and from this the authors of \cite{LowTtach} evaluated the finite temperature effective potential in the ideal gas limit. It is easy to show from this potential that at sufficiently low temperature the potential minimum shifts toward the origin of the complex $T$ plane, $T=0$, as the temperature increases.\footnote{We can show this in a similar way as we will perform in \S \ref{sec:lowT}.} On the other hand, computing the potential in the high temperature limit (using the standard method \cite{restoration}), we can show that the critical temperature, at which the potential has its minimum at $T=0$, is enormously higher than the Hagedorn temperature in the weak coupling limit. This is because the tension of the Dp-brane is very large if string coupling is very small as we can see from (\ref{eq:tension}), and tachyon field must have extremely high energy in order to create the {\D{p}} pair.

There is another approach to the finite temperature {\D{p}} system. Huang attempted to evaluate the finite temperature effective potential based on BSFT of bosonic strings \cite{Huang1}. He concluded that the high temperature tachyon potential has the same form with that of the zero temperature up to a temperature dependent string tension, and the tachyon rolls down towards the vacuum at $T= \infty$.\footnote{In a paper \cite{Huang2}, he investigated a finite temperature system of parallel brane and antibrane with a finite distance and concluded that it becomes stable at sufficiently high temperature.} However, even if we set $T=0$ in the potential, we cannot reproduce the free energy of on-shell open strings. Therefore, we need to compute the finite temperature effective potential in the presence of a constant tachyon background based on BSFT and statistical mechanics. Although we will focus on a coincident {\D{p}} pair in type II string theory, generalization to non BPS branes is straightforward.

In addition to the analogy of symmetry restoration of field theories, we have another motivation to study the finite temperature {\D{p}} system. The free energy of strings diverges above the Hagedorn temperature in general \cite{Hag}. However, in ten dimensional spacetime, we can reach the Hagedorn temperature for closed strings by supplying a finite amount of energy, while we need infinite energy to reach the Hagedorn temperature for open strings \cite{limiting}. From this it has been said that, in closed string case, the Hagedorn temperature is associated with a phase transition in analogy with the deconfining transition in QCD, while it is a `limiting temperature' in open string case. For this phase transition of closed strings, Sathiapalan \cite{Sa}, Kogan \cite{Ko} and Atick and Witten \cite{AW} have argued that the `winding modes' of Euclidean time direction becomes tachyonic above the Hagedorn temperature and a phase transition takes place due to the condensation of these tachyon fields. This phase transition is called the Hagedorn transition in closed string theory. However, so far we have not known where the minimum of this tachyon potential is and in what kind of backgrounds the system becomes stable. So we want to shed some light on this problem by investigating the finite temperature brane-antibrane system.

This paper is organized as follows. In \S \ref{sec:free} we give the free energy of open strings on the {\D{p}} system based on BSFT. In \S \ref{sec:lowT} we study the finite temperature effective potential at low temperature by using the canonical ensemble method. The system near the Hagedorn temperature is investigated in \S \ref{sec:comT} and \S \ref{sec:poten} by using the microcanonical ensemble method, because we cannot use the canonical ensemble method near the Hagedorn temperature as we will see in \S \ref{sec:comT}. In \S \ref{sec:Sdual}, we discuss a relationship between the Hagedorn transition of closed strings and the phase transition on the {\NS9} system. \S \ref{sec:conclusion} presents our conclusions and directions for future work. In appendix \ref{sec:canon}, we describe the finite temperature effective potential evaluated by using the canonical ensemble method.

\section{Free Energy of Open Strings}
\label{sec:free}

It is convenient to investigate the finite temperature effective potential by using the canonical ensemble method. In this method, the potential of the {\D{p}} system is given by the sum of the zero temperature tachyon potential and the free energy of open strings:
\BE
  V(T, \B) = 2 \tau_{p} \v \exp (-8 \T2) + F,
\label{eq:tacF}
\EE
where $F$ is the free energy of open strings. Thus, we begin by considering the free energy of open strings in this section.

We can compute the free energy by using Matsubara formalism. The free energy is given by the path integral of connected graphs of strings on the space where Euclidean time direction is compactified with the circumference of inverse temperature $\B$. Let us take the weak coupling approximation and treat strings as an ideal gas, that is, we ignore the interactions of open strings. We take into account only one-loop amplitude, where the corresponding cylinder world-sheets wind the Euclidean time direction at least once.

There have been attempts to generalize BSFT to the one-loop amplitude of open strings on the {\D{p}} system \cite{1loop}.\footnote{Strictly speaking, we must consider quantum master equation in order to investigate quantum corrections to the potential. However, the quantum master equation is still missing for BSFT. Thus, we will not discuss this subject.} However, there is an ambiguity in choosing the Weyl factors of the two boundaries of a one-loop world-sheet, because the conformal invariance is broken by the boundary terms in BSFT. Andreev and Oft have proposed the form of the one-loop amplitude of open strings on the {\D{p}} system \cite{1loopAO}. They deduced it from the principle that its low energy part should coincide with that of the tachyon field which obeys the action (\ref{eq:taceffaction1}). This amplitude can be obtained straightforwardly by choosing boundary terms as those of a cylinder world-sheet
\BE
  S_b = \int_{0}^{2 \pi t} d \tau \int_{0}^{\pi} d \sigma
    [ \T2 \delta (\sigma) + \T2 \delta (\pi - \sigma) ].
\EE
Here we restrict ourselves to the constant tachyon field, which we denote by $T$. The 1-loop amplitude in such background is then given by
\BA
  {\it Z_{1}} &=& \frac{16 \pi^4 i \v}{(2 \pi \ap)^{\frac{p+1}{2}}}
    \int_{0}^{\infty} \frac{d \tau}{\tau}
      (4 \pi \tau)^{- \frac{p+1}{2}} e^{-4 \pi \T2 \tau} \no \\
  && \times \left[
    \left( \frac{\th_3 (0 | i \tau)}{{\th_1}' (0 | i \tau)} \right)^4
      - \left(\frac{\th_2 (0 | i \tau)}{{\th_1}' (0 | i \tau)} \right)^4 \right].
\EA

We can obtain the same amplitude when we consider the field theory which has the mass spectra
\BA
  {M_{NS}}^2
    &=& \frac{1}{\ap} \left( N_B + N_{NS} + 2 \T2 - \frac{1}{2} \right),
\label{eq:massNS} \\
  {M_{R}}^2
    &=& \frac{1}{\ap} \left( N_B + N_{R} + 2 \T2 \right),
\label{eq:massR}
\EA
where $M_{NS}$ and $M_{R}$ are the mass of the Neveu-Schwarz and Ramond sectors, respectively, and $N_B$, $N_{NS}$ and $N_{R}$ are the oscillation modes of the boson, Neveu-Schwarz fermion and Ramond fermion, respectively. The lowest mode of the NS sector (\ref{eq:massNS}) coincides with that of the tachyon field (\ref{eq:tachyonmass}). It is convenient to express the free energy by the proper time form \cite{Pol} to compute the free energy from the above mass spectrum. It is given by
\BA
  F (\B) &=& - \frac{\v}{(2 \pi \ap)^{\frac{p+1}{2}}}
    \int_{0}^{\infty} \frac{d \tau}{\tau}
      (4 \pi \tau)^{- \frac{p+1}{2}} \sum_{{M_{NS}}^2} \sum_{r=1}^{\infty}
        \exp \left( -2 \pi \ap {M_{NS}}^2 \tau
          - \frac{r^2 \B^2}{8 \pi \ap \tau} \right) \no \\
  && + \frac{\v}{(2 \pi \ap)^{\frac{p+1}{2}}}
    \int_{0}^{\infty} \frac{d \tau}{\tau}
      (4 \pi \tau)^{- \frac{p+1}{2}} \sum_{{M_{R}}^2} \sum_{r=1}^{\infty}
        (-1)^r \exp \left( -2 \pi \ap {M_{R}}^2 \tau
          - \frac{r^2 \B^2}{8 \pi \ap \tau} \right).
\EA
\begin{figure}
\begin{center}
$${\epsfxsize=6.5 truecm \epsfbox{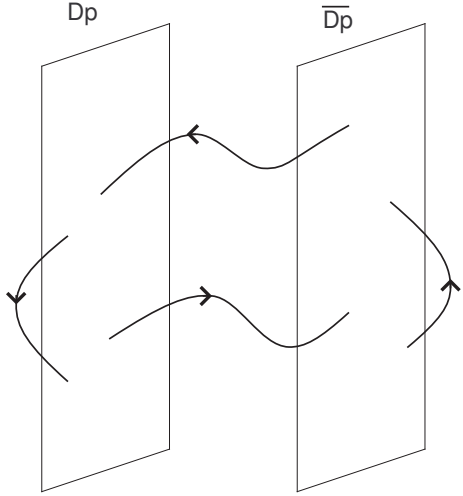}}$$
\caption{For clarity we have displayed the brane and antibrane to be separated. But we are considering the coincident brane-antibrane pair.}
\label{fig:D9}
\end{center}
\end{figure}
The difference in the signs of each terms come from the fact that the NS sector represents spacetime boson and the R sector spacetime fermion. As depicted in the Figure \ref{fig:D9}, there are four types of open strings, namely, two types of the strings whose two ends attach to the same brane and two types of the strings which are stretched between the Dp-brane and the $\overline{\textrm{Dp}}$-brane. If we denote the open strings going from a Dp-brane to a $\overline{\textrm{Dp}}$-brane as (Dp,$\overline{\textrm{Dp}}$), others are denoted as ($\overline{\textrm{Dp}}$,Dp), (Dp,Dp) and ($\overline{\textrm{Dp}}$,$\overline{\textrm{Dp}}$). We must impose the GSO projection for (Dp,Dp) and ($\overline{\textrm{Dp}}$,$\overline{\textrm{Dp}}$), and the opposite GSO projection for (Dp,$\overline{\textrm{Dp}}$) and ($\overline{\textrm{Dp}}$,Dp). So, instead of imposing GSO projection, we may multiply two as an overall factor. Summations over the mass can be rewritten by using $\th$ function as
\BA
  \sum_{{M_{NS}}^2} \exp \left( -2 \pi \ap {M_{NS}}^2 \tau \right)
    &=& 32 \pi^4 e^{-4 \pi \T2 \tau}
      \left( \frac{\th_3 (0 | i \tau)}{{\th_1}' (0 | i \tau)} \right)^4, \no \\
  \sum_{{M_{R}}^2} \exp \left( -2 \pi \ap {M_{R}}^2 \tau \right)
    &=& 32 \pi^4 e^{-4 \pi \T2 \tau}
      \left( \frac{\th_2 (0 | i \tau)}{{\th_1}' (0 | i \tau)} \right)^4.
\EA
Summations over $r$ can be rewritten by using $\th$ function as
\BA
  \sum_{r=1}^{\infty} \exp \left(- \frac{r^2 \B^2}{8 \pi \ap \tau} \right)
    &=& \frac{1}{2} \left[ \th_3 \left( 0 \left| \frac{i \B^2}{8 \pi^2 \ap \tau}
      \right. \right) -1 \right], \no \\
  \sum_{r=1}^{\infty} (-1)^r
    \exp \left(- \frac{r^2 \B^2}{8 \pi \ap \tau} \right)
      &=& \frac{1}{2}
        \left[ \th_4 \left( 0 \left| \frac{i \B^2}{8 \pi^2 \ap \tau}
          \right. \right) -1 \right].
\EA
Thus, we obtain\footnote{We can represent this free energy as the propagator of closed strings by using the modular transformation of $\th$ functions \cite{VM}.}
\BA
  F (T, \B) &=& - \frac{16 \pi^4 \v}{(2 \pi \ap)^{\frac{p+1}{2}}}
    \int_{0}^{\infty} \frac{d \tau}{\tau}
      (4 \pi \tau)^{- \frac{p+1}{2}} e^{-4 \pi \T2 \tau} \no \\
  && \hspace{2cm} \times \left[ \left(\frac{\th_3 (0 | i \tau)}
    {{\th_1}' (0 | i \tau)} \right)^4
      \left( \th_3 \left( 0 \left| \frac{i \B^2}{8 \pi^2 \ap \tau} \right.
        \right) -1 \right) \right. \no \\
  && \hspace{4cm} - \left.
    \left( \frac{\th_2 (0 | i \tau)}{{\th_1}' (0 | i \tau)} \right)^4
      \left( \th_4 \left( 0 \left| \frac{i \B^2}{8 \pi^2 \ap \tau} \right.
        \right) -1 \right) \right].
\label{eq:free}
\EA
We can reproduce the free energy of on-shell open strings on the {\D{p}} system by substituting $T=0$. We will discuss the finite temperature effective potential at low temperature based on this free energy in the next section.

\section{Finite Temperature Effective Potential at Low Temperature}
\label{sec:lowT}

As mentioned in \S \ref{sec:Intro}, the finite temperature effective potential of the tachyon field model is investigated by Danielsson, G\"{u}ijosa and Kruczenski \cite{LowTtach}. From their calculation we can find out that the potential minimum of the finite temperature tachyon potential shifts toward $T=0$ as the temperature increases. In this section, we will show that we can obtain the same result by calculating the finite temperature effective potential of the {\D{p}} system at low temperature.

In the large $\B$ limit, the integral in the free energy (\ref{eq:free}) can be approximated by the large $\tau$ part. This is because the $\th$ function depending on $\B$ can be expanded in
\BA
  \exp \left( - \frac{\B^2}{8 \pi \ap \tau} \right), \no
\EA
and these terms are suppressed exponentially unless $\tau$ is very large. Expanding $\th$ functions and extracting the leading term in large $\tau$ region, we obtain
\BA
  F (T, \B) \simeq - \frac{2 \v}{{\BH}^{p+1}}
    \int_{0}^{\infty} d \tau \ \tau^{\frac{p+3}{2}}
      \exp \left( - \pi (4 \T2 -1) \tau - \frac{\B^2}{8 \pi \ap \tau} \right),
        \no
\EA
where $\BH$ is the inverse of the Hagedorn temperature
\BE
  \BH = 2 \pi \sqrt{2 \ap}.
\EE
We should notice that this term corresponds to the contribution of the lowest energy (tachyonic) mode. Above equation can be rewritten by using the third kind of modified Bessel function
\BE
  F (T, \B) \simeq
    - 4 \v \left( \frac{\pi \sqrt{4 \T2 -1}}{\BH \B} \right)^{\frac{p+1}{2}}
      K_{\frac{p+1}{2}} \left( \frac{2 \pi \sqrt{4 \T2 -1} \ \B}{\BH} \right).
\EE
This free energy coincides with that of the tachyon field model when $p=9$ \cite{LowTtach}. If we assume that both $|T|$ and $\B$ are very large, the modified Bessel function can be approximated as
\BE
  K_{\nu} (z) \simeq \sqrt{\frac{\pi}{2z}} e^{-z},
\EE
and the free energy becomes
\BE
  F (T, \B) \simeq
    - \frac{2^{\frac{p}{2}+1} \pi^{\frac{p+1}{2}} \v |T|^{\frac{p}{2}}}
      {{\BH}^{\frac{p}{2}} \B^{\frac{p}{2}+1}}
        \exp \left( - \frac{4 \pi \B}{\BH} |T| \right).
\label{eq:lowTF}
\EE
We can obtain the finite temperature effective potential by substituting this free energy to (\ref{eq:tacF}). At the potential minimum, this potential satisfies
\BE
  \frac{ \p V(T, \B)}{\p |T|} = 0,
\EE
and we can obtain the condition for the temperature by substituting the potential we computed. Taking into consideration the assumption that both $|T|$ and $\B$ are very large, this gives
\BE
  |T| \simeq \frac{\pi}{2 \BH} \B = \frac{\pi {\cal T}_{\scriptscriptstyle H}}{2 {\cal T}},
\EE
where ${\cal T}$ denotes the temperature ${\cal T} = \B^{-1}$, and ${\cal T}_{\scriptscriptstyle H}$ denotes the Hagedorn temperature
\BE
  {\cal T}_{\scriptscriptstyle H} = {\BH}^{-1} = \frac{1}{2 \pi \sqrt{2 \ap}}.
\label{eq:HagedornT}
\EE

From this we can see that $|T|$ decreases with increasing ${\cal T}$. This implies that the potential minimum moves towards $|T|=0$ as the temperature increases, like the tachyon field model.

\section{Complex Temperature Formalism}
\label{sec:comT}

It is well-known that perturbative strings have a maximum temperature, which is so called the Hagedorn temperature. This comes from the fact that the degeneracy of oscillation modes of a single string increases exponentially with the energy, and the density of states $\Omega (E)$ of strings behave as
\BE
  \Omega (E) \sim e^{\BH E},
\EE
for large $E$. The energy of strings is dominated by very high oscillation modes of a single high energy string near the Hagedorn temperature. Consequently, we cannot raise the temperature by adding more energy to the system because a high energy string soaks up most of the energy.

Unfortunately, we cannot use the canonical ensemble method near the Hagedorn temperature \cite{Efura}. We can see this from the relation between density of states and partition function, that is, the Laplace transformation
\BE
  Z(\B) = \int_{0}^{\infty} dE \ \Omega (E) e^{- \B E},
\label{eq:Lap}
\EE
The integrand is not sharply peeked function of $E$ near the Hagedorn temperature, and the canonical ensemble method is not guaranteed to be valid. Thus, we must derive the thermodynamical quantities near the Hagedorn temperature from the microcanonical ensemble method, which is more fundamental than the canonical ensemble method in the sense that it is derived directly from ergodic theory. This process was performed in the case of closed strings \cite{Tan2} and in the case of open strings on D-branes \cite{Thermo}. We will apply this method in the case of open strings on the {\D{p}} system.

\begin{figure}
\begin{center}
$${\epsfxsize=6.5 truecm \epsfbox{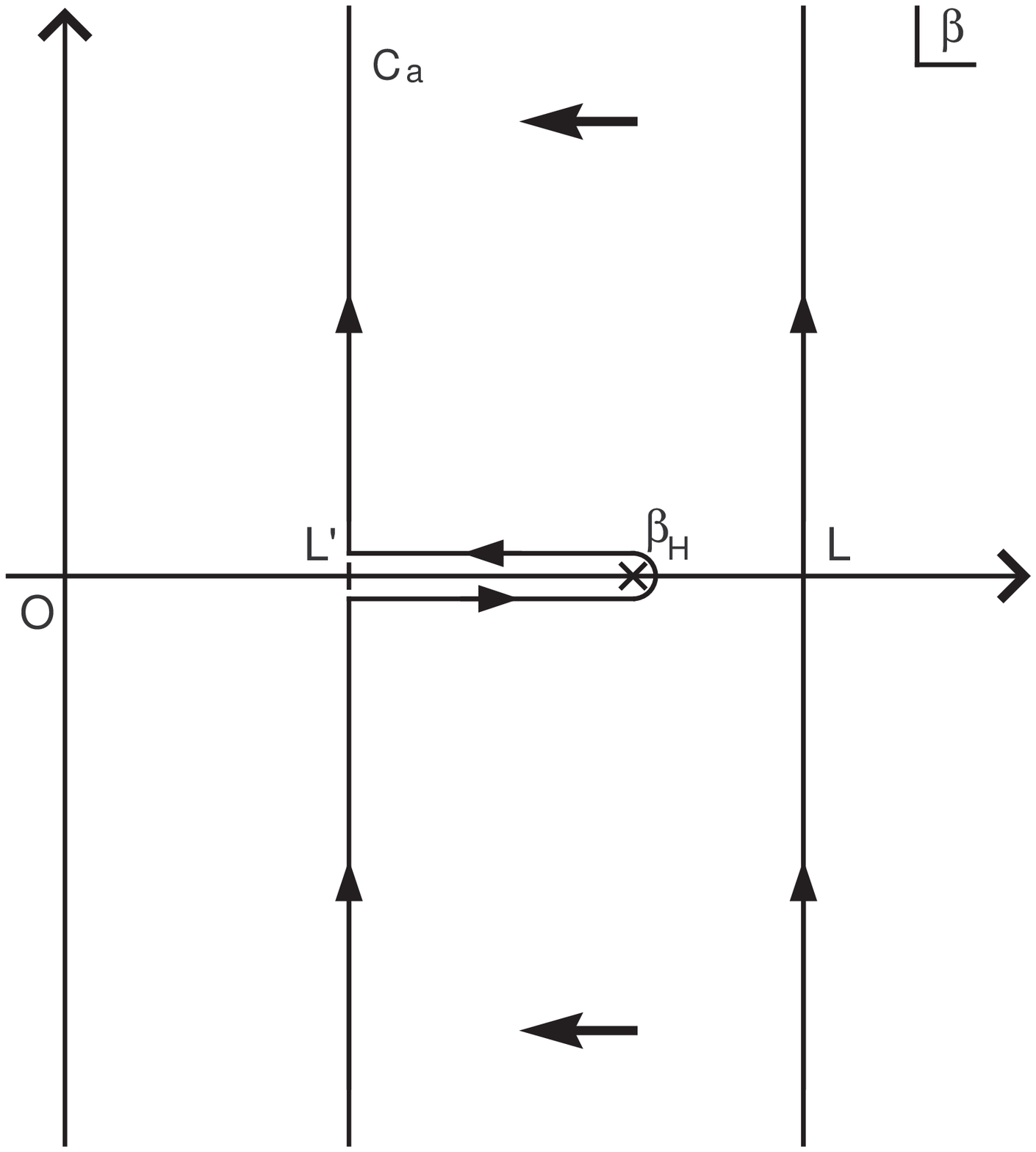}}$$
\caption{Complex $\beta$ plane.}
\label{fig:com1}
\end{center}
\end{figure}

We can derive all the statistical variables from the density of states $\Omega (E)$ in the microcanonical ensemble method. $\Omega (E)$ is given by the inverse Laplace transformation of $Z(\B)$
\BE
  \Omega (E) = \int_{L-i \infty}^{L+i \infty}
    \frac{d \B}{2 \pi i} Z(\B) e^{\B E},
\label{eq:inLap}
\EE
which is the opposite relation of (\ref{eq:Lap}). We must choose a constant $L$ such that the path of the integral in the complex $\B$-plane lies on the right side of all the singularities of the integrand. Then, we can deform the contour to the left side such as to pick picking up the singularities, as is sketched in Figure \ref{fig:com1}. We will first investigate the singular property of the partition function on the complex $\B$ plane in this section, and then calculate the finite temperature effective potential in the next section.

A general property of the partition function of strings is that it has the leading singularity at $\B = \BH$, which we call the Hagedorn singularity. In order to see this, let us see the behavior of the free energy near the Hagedorn temperature. In contrast to the low temperature case, we must expand the integral in small $\tau$ region. This can be easily performed by making the variable transformation
\BA
  \tau = \frac{1}{t}, \no
\EA
and considering large $t$ region. Using the modular transformation of $\th$ functions, we obtain
\BA
  F (T, \B) &=& - \frac{16 \pi^4 \v}{{\BH}^{p+1}}
    \int_{0}^{\infty}
      dt \ t^{\frac{p-9}{2}} \exp \left( - \frac{4 \pi \T2}{t} \right) \no \\
  && \hspace{2cm}
    \times \left[ \left(\frac{\th_3 (0 | i t)}{{\th_1}' (0 | it)} \right)^4
      \left( \th_3 \left( 0 \left| \frac{i \B^2 t}{8 \pi^2 \ap} \right. \right)
        -1 \right) \right. \no \\
  && \hspace{4cm} \left.
    - \left(\frac{\th_4 (0 | it)}{{\th_1}' (0 | it)} \right)^4
      \left( \th_4 \left( 0 \left| \frac{i \B^2 t}{8 \pi^2 \ap} \right. \right)
        -1 \right) \right].
\label{eq:Fmodular}
\EA
Expanding $\th$ functions and extracting the leading term in large $t$ region near the Hagedorn singularity, we obtain\footnote{Strictly speaking, we must introduce a low energy cutoff as we will see in appendix A.}
\BE
  F (T, \B) \simeq - \frac{4 \v}{{\BH}^{p+1}}
    \int_{0}^{\infty} dt
      \ t^{\frac{p-9}{2}} \exp \left[ - \frac{4 \pi \T2}{t}
        + \left( \pi - \frac{\B^2}{8 \pi \ap} \right) t \right].
\label{eq:FHag}
\EE
Since we will compute the $\T2$ term of the finite temperature effective potential in the vicinity of $T=0$, let us expand the free energy in $\T2$ and keep the lower order terms
\BA
  F (T, \B) &\simeq& - \frac{4 \v}{{\BH}^{p+1}} \int_{0}^{\infty} dt
    \ t^{\frac{p-9}{2}}
      \exp \left( \pi \frac{{\BH}^2 - \B^2}{{\BH}^2} t \right) \no \\
  &&+ \frac{16 \pi \v \T2}{{\BH}^{p+1}} \int_{0}^{\infty} dt
    \ t^{\frac{p-11}{2}}
      \exp \left( \pi \frac{{\BH}^2 - \B^2}{{\BH}^2} t \right).
\label{eq:highTF}
\EA
From this we will discuss the finite temperature effective action by using the canonical ensemble method in appendix A. If $p \neq 9$, the integration over $t$ can be easily done by using the formula:
\BE
  t^l = \frac{1}{\Gamma (-l)} \int_{0}^{\infty}
    du \ u^{- (l+1)} e^{-tu},
\EE
and we obtain
\BE
  F (T, \B) \simeq - \frac{4 \v}{{\BH}^{p+1}
    \Gamma (\frac{9-p}{2})} \int_{0}^{\infty} du
      \frac{u^{\frac{7-p}{2}}}{u- \pi \frac{{\BH}^2 - \B^2}{{\BH}^2}}
        + \frac{16 \pi \v \T2}{{\BH}^{p+1} \Gamma (\frac{11-p}{2})}
          \int_{0}^{\infty} du
            \frac{u^{\frac{9-p}{2}}}{u- \pi \frac{{\BH}^2 - \B^2}{{\BH}^2}}.
\EE
In order to investigate the behavior of the partition function, let us introduce
\BE
  W(T, \B) \equiv \log Z = - \B F,
\label{eq:Wdef}
\EE
and investigate the singular part of $W$, which we denote $W_{sing}$. $W$ has a branch point at $\B = \BH$, and we can draw a cut to the negative direction of the real axis. Let us evaluate the discontinuity of $W$ across the cut. From the definition of the discontinuity, it can be written as 
\BA
  \Delta W &\equiv& \lim_{\varepsilon \rightarrow +0}
    \left[ W(T, \B +i \varepsilon) - W(T, \B -i \varepsilon) \right]
      \ \ \ \ \ \ (\B < \BH) \no \\
  &\simeq& - \frac{4 \v}{{\BH}^{7-2 \a} \Gamma (\a +1)}
    \int_{0}^{\infty} du \ u^{\a} \lim_{\varepsilon' \rightarrow +0}
      \left( \frac{1}{u- 2 \pi \frac{\BH - \B}{\BH} +i \varepsilon'}
        - \frac{1}{u- 2 \pi \frac{\BH - \B}{\BH} -i \varepsilon'} \right) \no \\
  &&+ \frac{16 \pi \v \T2}{{\BH}^{7-2 \a} \Gamma (\a +2)}
    \int_{0}^{\infty} du \ u^{\a +1} \lim_{\varepsilon' \rightarrow +0}
      \left( \frac{1}{u- 2 \pi \frac{\BH - \B}{\BH} +i \varepsilon'}
        - \frac{1}{u- 2 \pi \frac{\BH - \B}{\BH} -i \varepsilon'} \right), \no
\EA
where $\varepsilon' = 2 \pi \varepsilon / \BH$ and we have introduced
\BE
  \a \equiv \frac{7-p}{2}.
\EE
By using the definition of $\delta$ function
\BE
  \delta (x) \equiv \frac{1}{2 \pi i} \lim_{\varepsilon \rightarrow +0}
    \left( \frac{1}{x-i \varepsilon} - \frac{1}{x+i \varepsilon} \right),
\EE
we can rewrite $\Delta W$ as
\BE
  \Delta W \simeq - \frac{(-1)^{\a + \frac{1}{2}}
    4 (2 \pi)^{\a +1} \v}{\Gamma (\a +1) {\BH}^{7- \a}} (\B - \BH)^{\a}
      + \frac{(-1)^{\a + \frac{3}{2}}
        8 (2 \pi)^{\a +3} \v \T2}{\Gamma (\a +2) {\BH}^{8- \a}}
          (\B - \BH)^{\a +1}.
\EE
This non-zero $\Delta W$ guarantees that $\B = \BH$ is a branch point. We can deduce $W_{sing}$ from $\Delta W$ as follows \cite{Tan2} \cite{Thermo}.

\renewcommand{\descriptionlabel}[1]{\large\bfseries{#1}}
\begin{description}

\item[(a)] {\large \bf \ $\a :$ half-integer ($p :$ even)}
\vspace{0.5cm}

$(\B - \BH)^{\a}$ has discontinuity at $\B = \BH$ when $\a$ is a half-integer. Using this fact and a formula for $\Gamma$ function
\BE
  \Gamma (Z) \Gamma (1-z) = \frac{\pi}{\sin \pi z},
\label{eq:GGsin}
\EE
we can express $W_{sing}$ as
\BE
  W_{sing} \sim [C_{\a} (\B - \BH)^{\a}
    - D_{\a} \T2 (\B - \BH)^{\a +1}] \v,
\EE
where
\BA
  C_{\a} &=& \frac{4 (2 \pi)^{\a} \Gamma (- \a )}{{\BH}^{7- \a}}, \\
  D_{\a} &=& \frac{8 (2 \pi)^{\a +2} \Gamma (- \a -1)}{{\BH}^{8- \a}}.
\EA

\item[(b)] {\large \bf $\a :$ integer ($\a \neq -1$) ($p :$ odd ($p \neq 9$))}
\vspace{0.5cm}

In this case, $(\B - \BH)^{\a}$ has no discontinuity. However, $\log [{\BH}^{-1} (\B - \BH)]$  has discontinuity at $\B = \BH$ and we can express $W_{sing}$ as
\BE
  W_{sing} \sim [C_{\a} (\B - \BH)^{\a}
    - D_{\a} \T2 (\B - \BH)^{\a +1}] \v \log \left( \frac{\B - \BH}{\BH} \right),
\EE
where
\BA
  C_{\a} &=& \frac{(-1)^{\a +1} 4 (2 \pi)^{\a}}{\Gamma (\a +1) {\BH}^{7- \a}}, \\
  D_{\a} &=& \frac{(-1)^{\a +2} 8 (2 \pi)^{\a +2}}{\Gamma (\a +2) {\BH}^{8- \a}}.
\EA

\item[(c)] {\large \bf $\a = -1$ ($p=9$)}
\vspace{0.5cm}

In this case, we can perform the integral in the first term of (\ref{eq:highTF}) explicitly. For the second term, we can obtain the discontinuity in the same way when $\a$ is an integer. Combining two terms, we can express $W_{sing}$ as
\BE
  W_{sing} \sim \frac{\Cm \v}{\B - \BH}
    - \Dm \v \T2 \log \left( \frac{\B - \BH}{\BH} \right),
\label{eq:Wsingp9}
\EE
where
\BA
  \Cm &=& \frac{2}{\pi {\BH}^8}, \\
  \Dm &=& - \frac{16 \pi}{{\BH}^9}.
\EA
The first term gives the pole in the complex $\B$-plane.

\end{description}

We thus have obtained $W_{sing}$ for all possible $\a$, which can summarized as
\BE
  W_{sing} \sim C_{\a} \v (\B - \BH)^{\a}
    \left( \log \left( \frac{\B - \BH}{\BH} \right) \right)^a
      -D_{\a} \v \T2 (\B - \BH)^{\a} 
        \left( \log \left( \frac{\B - \BH}{\BH} \right) \right)^b,
\label{eq:Wsing}
\EE
where $a$ is $1$ if $\a$ is an integer except $-1$ and $0$ otherwise, and $b$ is $1$ if $\a$ is an integer and $0$ otherwise. The singular part of the partition function can be obtained from (\ref{eq:Wdef}) and this $W_{sing}$. We can compute the density of states from it. We will see it in the next section.

\section{Finite Temperature Effective Potential near the Hagedorn Temperature}
\label{sec:poten}

We can derive thermodynamical variables from the density of states in the microcanonical ensemble method, as we mentioned in the previous section. The density of states $\Omega (E)$ can be obtained from the partition function $Z(\B)$ by using the inverse Laplace transformation (\ref{eq:inLap}). $Z(\B)$ is related to $W(\B)$ as
\BE
  Z(T, \B) = e^W = \exp [W_{reg} + W_{sing}].
\label{eq:ZeW}
\EE
Here $W_{reg}$ is the regular part of $W$, which can be expanded in a power of $(\B - \BH)$ as
\BE
  W_{reg} = \lambda_0 \v - \sigma_0 \v (\B - \BH) + O( \v (\B - \BH)^2).
\label{eq:Wreg}
\EE
It is easily to see that $\lambda_0$ is the constant which has the dimension of the inverse of volume, and $\sigma_0$ is the constant which has the dimension of an energy density. Thus, using eqs. (\ref{eq:inLap}), (\ref{eq:Wsing}), (\ref{eq:ZeW}) and (\ref{eq:Wreg}), we obtain the density of states as
\BA
  \Omega (T,E) &\simeq& e^{\BH E + \lambda_0 \v}
    \left. \int_{C_a} \frac{d \B}{2 \pi i}
      \exp \right[ (\B - \BH) \Eb + O( \v (\B - \BH)^2) \no \\
  && \hspace{4cm} +C_{\a} \v (\B - \BH)^{\a}
    \left( \log \left( \frac{\B - \BH}{\BH} \right) \right)^a \no \\
  && \hspace{4cm} \left. -D_{\a} \v \T2 (\B - \BH)^{\a} 
        \left( \log \left( \frac{\B - \BH}{\BH} \right) \right)^b \right],
\label{eq:dos}
\EA
where $\Eb \equiv E - \sigma_0 \v$. Furthermore, knowing the explicit form of $\Omega (E)$, we can calculate the entropy $S(E)$ as
\BE
  S(T,E) = \log \Omega (T,E) \delta E,
\label{eq:Sdef}
\EE
and the inverse temperature $\B$ as
\BE
  \B = \frac{\p S}{\p E}.
\label{eq:Bdef}
\EE
Then, the finite temperature effective potential of the {\D{p}} system is given by
\BE
  V(T,E) = 2 \tau_{p} \v \exp (-8 \T2) - \B^{-1} S.
\label{eq:Vdef}
\EE
Now, we are ready to calculate the finite temperature effective potential by using the microcanonical ensemble method. We will give a discussion for each value of $p$ separately, since our approximation to calculate $\Omega (T,E)$ differs for each value of $p$.

\renewcommand{\descriptionlabel}[1]{\large\bfseries{#1}}
\begin{description}

\item[(a)] {\large \bf \ $p=9$  ($\a = -1$)}
\vspace{0.5cm}

Let us first consider the case of the {\D{9}} system, which is the most interesting case. In this case (\ref{eq:dos}) is rewritten as
\BE
  \Omega (T,E) \simeq e^{\BH E + \lambda_0 \v} \int_{C_a} \frac{d \B}{2 \pi i}
    \left( \frac{\B - \BH}{\BH} \right)^{- \Dm \v \T2}
      \exp \left[ (\B - \BH) \Eb + \frac{\Cm \v}{\B - \BH} \right].
\label{eq:p9dos}
\EE
Let us suppose that both $E$ and $E / \v$ are very large. Then, the saddle point method works well because the exponent in the integrand is very large. The result is\footnote{We can obtain the same result by using modified Bessel function.}
\BE
  \Omega (T,E) \simeq \frac{1}{2 \sqrt{\pi}}
    \left( \frac{{\BH}^2 \Eb}{\Cm \v}
      \right)^{\frac{1}{2} \Dm \v \T2}
        \left( \frac{\Cm \v}{\Eb^3} \right)^{\frac{1}{4}}
          \exp \left( \BH E + \lambda_0 \v
            +2 \sqrt{\Cm \v \Eb} \right).
\EE
In (\ref{eq:p9dos}), we ignored $O( \v (\B - \BH)^2)$ term when we derive this equation from (\ref{eq:dos}). We will examine the correction of the saddle point method from this term later in the $p=6$ case. Substituting $\Omega (E)$ into (\ref{eq:Sdef}), we obtain
\BA
  S(T,E) \simeq \frac{\Dm \v \T2}{2} \log \left( \frac{{\BH}^2 \Eb}{\Cm \v} \right)
    - \frac{3}{4} \log \left( \frac{\Eb}{{\Cm}^{\frac{1}{3}}
      \v^{\frac{1}{3}} (\delta E)^{\frac{4}{3}}} \right) \no \\
        + \BH E + \lambda_0 \v + 2 \sqrt{\Cm \v \Eb},
\EA
and from (\ref{eq:Bdef}), we get
\BE
  \B \simeq \frac{\Dm \v \T2}{2 \Eb}
    - \frac{3}{4 \Eb} + \BH + \sqrt{\frac{\Cm \v}{\Eb}}.
\label{eq:Bmin}
\EE
The finite temperature effective potential can be derived from (\ref{eq:Vdef}). In order to argue the stability of the brane-antibrane system, we need only 
$\T2$ term of $V(T,E)$. This term is given by
\BE
  \left[ -16 \tau_9 \v
   + \frac{8 \pi \v}{{\BH}^{10}}
    \log \left( \frac{\pi {\BH}^{10} \Eb}{2 \v}
     \right) \right] \T2.
\label{eq:p9T2E}
\EE
It should be noted that the second term in the coefficient of $\T2$ increases with increasing $E$. Since the first term is constant as far as $\v$ and $\tau_9$ fixed, the sign of the $\T2$ term changes from negative to positive at large $E$. The coefficient vanishes when
\BE
  \Eb \simeq \frac{2 \v}{\pi {\BH}^{10}} \exp \left( \frac{2 {\BH}^{10} \tau_9}{\pi} \right).
\EE
If we approximate (\ref{eq:Bmin}) as
\BE
  \B \simeq \BH + \sqrt{\frac{\Cm \v}{\Eb}},
\EE
we can derive the critical temperature ${\cal T}_c$ at which the coefficient vanishes. The result is
\BE
  {\cal T}_c = \beta^{-1}
   \simeq {\BH}^{-1}
    \left[ 1+ \exp \left( - \frac{{\BH}^{10} \tau_9}
     {\pi} \right) \right]^{-1}.
\label{eq:Tc}
\EE
Here we see that this temperature is very close to the Hagedorn temperature since $\tau_9$ is very large if the coupling of strings is very small. Above this temperature, the coefficient of $\T2$ is positive and $T=0$ becomes the potential minimum. This implies that a phase transition occurs at the temperature ${\cal T}_c$ which is slightly below the Hagedorn temperature, and the {\D{9}} system is stable above this temperature.

\vspace{0.5cm}
\item[(b)] {\large \bf \ $p=8$  ($\a = - 1/2$)}
\vspace{0.5cm}

Next, let us consider the case of the {\D{8}} system, where we will see that a result strikingly different from the case of the {\D{9}} system arises. From (\ref{eq:dos}) the density of states is given by
\BA
  \Omega (T,E) &\simeq& e^{\BH E + \lambda_0 \v} \int_{C_a} \frac{d \B}{2 \pi i}
    \exp \left[ (\B - \BH) \Eb + \Ch \v (\B - \BH)^{- \frac{1}{2}}
      - \Dh \v \T2 (\B - \BH)^{\frac{1}{2}} \right] \no \\
  &\simeq& e^{\BH E + \lambda_0 \v} \int_{C_a} \frac{d \B}{2 \pi i}
    \left[ 1- \Dh \v \T2 (\B - \BH)^{\frac{1}{2}} \right] \no \\
      && \hspace{4cm} \times \exp \left[ (\B - \BH) \Eb
        + \Ch \v (\B - \BH)^{- \frac{1}{2}} \right],
\EA
where we take the small $\T2$ approximation in the second equality. We can also use the saddle point method and obtain
\BE
  \Omega (T,E) \simeq \frac{{\Ch}^{\frac{1}{3}} {\v}^{\frac{1}{3}}}
    {3^{\frac{1}{2}} 2^{\frac{1}{3}} \pi^{\frac{1}{2}} \Eb^{\frac{5}{6}}}
      \exp \left( \BH E + \lambda_0 \v
        + \frac{3 {\Ch}^{\frac{2}{3}} \v^{\frac{2}{3}} \Eb^{\frac{1}{3}}}
          {2^{\frac{2}{3}}}
            - \frac{{\Ch}^{\frac{1}{3}} \Dh \v^{\frac{4}{3}} \T2}
              {2^{\frac{1}{3}} \Eb^{\frac{1}{3}}} \right).
\EE
We can calculate the entropy $S$, the inverse temperature $\B$ and the potential $V(T,E)$ from (\ref{eq:Sdef}), (\ref{eq:Bdef}) and (\ref{eq:Vdef}) as in the {\D{9}} case. The $\T2$ term of $V(T,E)$ is given by
\BE
  \left[ -16 \tau_8 \v
   - \frac{2^{\frac{23}{3}} \v^{\frac{4}{3}}}{3 {\BH}^{12} \Eb^{\frac{1}{3}}}
     \right] \T2.
\label{eq:T2p8}
\EE
It should be noted that the second term in the coefficient of $\T2$ decreases as $E$ gets large. Thus, the coefficient of $\T2$ remains negative for large $E$. This implies that a phase transition does not occur unlike in the {\D{9}} case.

\vspace{0.5cm}
\item[(c)] {\large \bf \ $p=7$  ($\a = 0$)}
\vspace{0.5cm}

In this case, the density of states can be obtained from (\ref{eq:dos}) as
\BE
  \Omega (T,E) \simeq e^{\BH E + \lambda_0 \v} \int_{C_a} \frac{d \B}{2 \pi i}
    \exp \left[ (\B - \BH) \Eb + \left( C_0 - D_0 \T2 (\B - \BH) \right)
      \v \log \left( \frac{\B - \BH}{\BH} \right) \right],
\EE
and we can also use the saddle point method if we expand in $\T2$ like in the {\D{8}} case. The result is
\BE
  \Omega (T,E) \simeq \frac{\sqrt{-C_0 \v}}{\sqrt{2 \pi} \Eb}
    \exp \left[ \BH E + \lambda_0 \v -C_0 \v
      + \left( C_0 \v + \frac{C_0 D_0 \v^2 \T2}{\Eb} \right)
        \log \left( - \frac{C_0 \v}{\BH \Eb} \right) \right].
\EE
We can calculate the finite temperature effective potential from this, and its $\T2$ term is given by
\BE
  \left[ -16 \tau_7 \v
   - \frac{2^7 \pi^2 \v^2}{{\BH}^{16} \Eb}
     \log \left( \frac{{\BH}^8 \Eb}{4 \v} \right) \right] \T2.
\label{eq:T2p7}
\EE
From this we can see that the coefficient is always negative, so that a phase transition does not occur.

\vspace{0.5cm}
\item[(d)] {\large \bf \ $p=6$  ($\a = 1/2$)}
\vspace{0.5cm}

In this case, the density of states can be obtained from (\ref{eq:dos}) as
\BE
  \Omega (T,E) \simeq e^{\BH E + \lambda_0 \v} \int_{C_a} \frac{d \B}{2 \pi i}
    \exp \left[ (\B - \BH) \Eb
      + \left( C_{\frac{1}{2}} (\B - \BH)^{\frac{1}{2}}
        - D_{\frac{1}{2}} \T2 (\B - \BH)^{\frac{3}{2}} \right) \v \right]
\EE
and we can also use the saddle point method if we expand in $\T2$ like in the {\D{8}} case. The result is
\BE
  \Omega (T,E) \simeq - \frac{C_{\frac{1}{2}} \v}{2 \sqrt{\pi} \Eb^{\frac{3}{2}}}
    \exp \left[ \BH E + \lambda_0 \v
      - \frac{{C_{\frac{1}{2}}}^2 \v^2}{4 \Eb}
        + \frac{{C_{\frac{1}{2}}}^3 D_{\frac{1}{2}} \v^4}{8 \Eb^3} \T2 \right]
\EE
We can calculate the finite temperature effective potential from this, and its $\T2$ term is given by
\BE
  \left[ -16 \tau_6 \v
   + \frac{2^{17} \pi^6 \v^4}{{\BH}^{28} \Eb^3} \right] \T2
\EE
From this we can see that the coefficient is always negative, so that a phase transition does not occur.

Here, we must consider corrections to the saddle point approximation. Such corrections come from $O( \v (\B - \BH)^2)$ term in (\ref{eq:dos}), which we have ignored so far. This term gives rise to a shift of the saddle point. Let us consider in the case that $\a$ is a half integer. If we write this term as
\BE
  \sum_{n=0}^{\infty} O( \v (\B - \BH)^{n+2}),
\EE
then the correction enters as a multiplicative factor
\BE
  \exp \left[ \sum_{n=0}^{\infty}
    O \left( \v \left( \frac{\Eb}{\v} \right)^{\frac{n+2}{\a -1}} \right)
      \right].
\label{eq:T2p6}
\EE
If the ratio $\Eb / \v$ is very large, we can use the saddle point method for $\a \leq 1/2$, because the power of $\Eb / \v$ is negative and the correction becomes negligible. However, we cannot ignore the correction for $\a \geq 3/2$, because the power is positive and the correction becomes very large for sufficiently large $n$. Similar calculation leads us to the result that we can use the saddle point method also in the case of $\a = 0$, $-1$. If $\a$ is a positive integer then $O( \v (\B - \BH)^2)$ term is of the same order to other terms. Therefore, we cannot use the saddle point method when $\a \geq 1$.

\vspace{0.5cm}
\item[(e)] {\large \bf \ $p=5$  ($\a = 1$)}
\vspace{0.5cm}

In this case, we cannot use the saddle point method, and thus we adapt another approximation as follows. The density of states can be obtained from (\ref{eq:dos}) as
\BA
  \Omega (T,E) &\simeq& e^{\BH E + \lambda_0 \v}
    \left. \int_{C_a} \frac{d \B}{2 \pi i}
      \exp \right[ (\B - \BH) \Eb \no \\
  && \left. + \left( C_1 (\B - \BH) - D_1 \T2 (\B - \BH)^2 \right)
    \v \log \left( \frac{\B - \BH}{\BH} \right) \right].
\label{eq:dosp5}
\EA
If we define
\BE
  z = - (\B - \BH) E',
\EE
then (\ref{eq:dosp5}) can be rewritten as
\BA
  \Omega (T,E) &\simeq& - \frac{e^{\BH E + \lambda_0 \v}}{E'}
    \int_{C_b} \frac{dz}{2 \pi i} e^{-z} \no \\
  && \hspace{2cm} \times \exp \left[ - \frac{C_1 \v z}{E'}
    \left( \log \left( \frac{\v z}{{\BH}^6 E'} \right) - \pi i \right)
      \right. \no \\
  && \hspace{4cm} \left. - \frac{D_1 \v \T2 z^2}{{E'}^2}
    \left( \log \left( \frac{z}{\BH E'} \right) - \pi i \right) \right],
\label{eq:p5dos}
\EA
where we define
\BE
  E' \equiv \Eb - C_1 \v \log \left( \frac{\v}{{\BH}^5} \right),
\EE
for simplicity. The contour of the integral is deformed in the complex plane, as sketched in Figure \ref{fig:com2}. We approximate the integral in (\ref{eq:p5dos}) by the contour integral in both side of the cut as
\BA
  && \int_{z_{1}}^{0} \frac{dz}{2 \pi i} e^{-z}
    \exp \left[ - \frac{C_1 \v z}{E'}
      \left( \log \left( \frac{\v z}{{\BH}^6 E'} \right) - \pi i \right)
        - \frac{D_1 \v \T2 z^2}{{E'}^2}
          \left( \log \left( \frac{z}{\BH E'} \right) - \pi i \right)
            \right] \no \\
  && + \int_{0}^{z_{1}} \frac{dz}{2 \pi i} e^{-z}
    \exp \left[ - \frac{C_1 \v z}{E'}
      \left( \log \left( \frac{\v z}{{\BH}^6 E'} \right) + \pi i \right)
        - \frac{D_1 \v \T2 z^2}{{E'}^2}
          \left( \log \left( \frac{z}{\BH E'} \right) + \pi i \right)
            \right] \no \\
  && = \int_{0}^{z_{1}} \frac{dz}{\pi} e^{-z}
    \exp \left[ - \frac{C_1 \v z}{E'}
      \log \left( \frac{\v z}{{\BH}^6 E'} \right)
        - \frac{D_1 \v \T2 z^2}{E'^2}
          \log \left( \frac{z}{\BH E'} \right) \right] \no \\
  && \hspace{2cm} \times \sin \left( \frac{\pi C_1 \v z}{E'}
              + \frac{\pi D_1 \v \T2 z^2}{{E'}^2} \right), \no
\EA
where
\BE
  z_{1} = - (L' - \BH) E'.
\EE
Since we are assuming that $E$ is very large, the integrand can be expanded in $1/ E'$, and we can take a limit $z_{1} \rightarrow \infty$. The integral becomes
\BE
  - \frac{C_1 \v}{E'} \Gamma (2)
    + \left[ \frac{{C_1}^2 \v^2}{{E'}^2 }
      \log \left( \frac{\v}{{\BH}^6 E'} \right)
        - \frac{D_1 \v \T2}{{E'}^2} \right] \Gamma (3)
          + \frac{{C_1}^2 \v^2}{{E'}^2 } \Gamma' (3),
\EE
where we express the derivative of $\Gamma$ function as $\Gamma' (z)$. If we include $O( \v (\B - \BH)^2)$ term in (\ref{eq:dos}), it can be rewritten as $O( \v z^2 / {E'}^2)$ and is a function of $(\B - \BH)$ to the integer power, so that we can drop it in this expansion. We can calculate $\Gamma' (z)$ from the relation between $\Gamma$ function and digamma function $\psi (z)$
\BE
  \psi (z) = \frac{\Gamma' (z)}{\Gamma (z)}.
\EE
and from
\BE
  \psi (3) = \frac{3}{2} - \gamma,
\EE
where $\gamma$ is the Euler constant $\gamma =0.57721 \cdot \cdot \cdot$. Substituting the result into (\ref{eq:p5dos}), we obtain
\BE
  \Omega (T,E) \simeq \frac{8 \pi \v}{{\BH}^6 {E'}^2}
    e^{\BH E + \lambda_0 \v}
      \left[ 1- \frac{16 \pi \v}{{\BH}^6 E'}
        \log \left( \frac{\v}{{\BH}^6 E'} \right)
          - \frac{8 \pi^2 \T2}{\BH E'}
            - \frac{8 \pi (3-2 \gamma) \v}{{\BH}^6 E'} \right].
\EE
We can calculate the finite temperature effective potential from this, and its $\T2$ term is given by
\BE
  \left[ -16 \tau_5 \v
   + \frac{16 \pi^2}{{\BH}^{2} E'} \right] \T2.
\EE
From this we can see that the coefficient is always negative, so that a phase transition does not occur.
\begin{figure}
\begin{center}
$${\epsfxsize=6.5 truecm \epsfbox{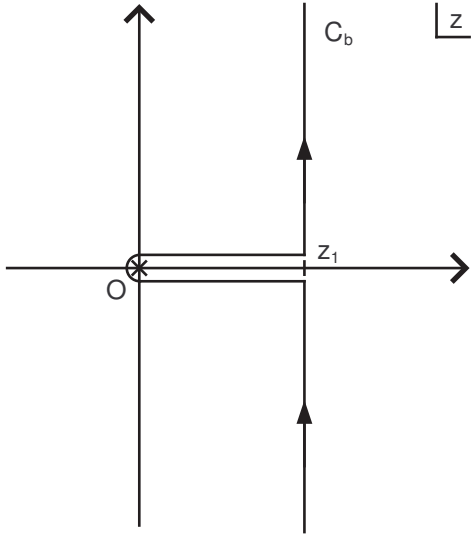}}$$
\caption{Complex $z$ plane.}
\label{fig:com2}
\end{center}
\end{figure}

\vspace{0.5cm}
\item[(f)] {\large \bf \ $p=4,2,0$  ($\a = 3/2,5/2,7/2$)}
\vspace{0.5cm}

We can calculate the finite temperature effective potential in analogy with the $p=5$ case. The density of states can be obtained from (\ref{eq:dos}) as
\BE
  \Omega (T,E) \simeq e^{\BH E + \lambda_0 \v} \int_{C_a} \frac{d \B}{2 \pi i}
    \exp \left[ (\B - \BH) \Eb
      + \left( C_{\a} (\B - \BH)^{\a}
        - D_{\a} \T2 (\B - \BH)^{\a +1} \right) \v \right].
\EE
Using the transformation
\BE
  z=-(\B - \BH) \Eb,
\EE
we get
\BE
  \Omega (T,E) \simeq - \frac{e^{\BH E + \lambda_0 \v}}{\Eb}
    \int_{C_b} \frac{dz}{2 \pi i}
      \exp \left[ -z
        + \left( \frac{C_{\a} z^{\a}}{\Eb^{\a}} e^{- \pi i \a}
          - \frac{D_{\a} \T2 z^{\a +1}}{\Eb^{\a +1}} e^{- \pi i (\a +1)} \right) \v \right].
\EE
The contour of the integral is deformed as in Figure \ref{fig:com2}, but $z_1$ is here replaced by $z_2$, where
\BE
  z_{2} =-(L' - \BH) \Eb.
\EE
We can approximate the integral by the contour integral in both sides of the cut as
\BE
  - \int_{0}^{z_{2}} \frac{dz}{\pi} e^{-z}
    \sin \left( \frac{C_{\a} \sin (- \pi \a) \v z^{\a}}{\Eb^{\a}}
      - \frac{D_{\a} \sin (- \pi (\a +1)) \v \T2 z^{\a +1}}
        {\Eb^{\a +1}} \right).
\EE
Expanding the integrand in $1/ \Eb$ and taking the limit $z_{2} \rightarrow \infty$, the integral becomes
\BA
  &&- \frac{4 (2 \pi)^{\a} \Gamma (- \a) \Gamma (\a +1) \sin (- \pi \a)}
    {\pi {\BH}^{7- \a}}
      \frac{\v}{\Eb^{\a}} \no \\
  && \hspace{1cm} + \frac{8 (2 \pi)^{\a +2} \Gamma (- \a -1) \Gamma (\a +2)
    \sin (- \pi (\a +1)) \T2}{\pi {\BH}^{8- \a}}
      \frac{\v}{\Eb^{\a +1}}.
\EA
Using (\ref{eq:GGsin}), we obtain
\BE
  \Omega (T,E) \simeq - \frac{4 (2 \pi)^{\a} \v}{{\BH}^{7- \a} \Eb^{\a +1}}
    e^{\BH E + \lambda_0 \v}
      \left( 1- \frac{8 \pi^2 \T2}{\BH \Eb} \right).
\EE
We can calculate the finite temperature effective potential from this, and its $\T2$ term is given by
\BE
  \left[ -16 \tau_p \v
   + \frac{16 \pi^2}{{\BH}^{2} \Eb} \right] \T2.
\EE
From this we can see that the coefficient is always negative, so that a phase transition does not occur.

\vspace{0.5cm}
\item[(g)] {\large \bf \ $p=3,1$  ($\a = 2,3$)}
\vspace{0.5cm}

We can also calculate the finite temperature effective potential in a similar way as that we have done in the above two cases. The density of states can be obtained from (\ref{eq:dos}) as
\BA
  \Omega (T,E) &\simeq& e^{\BH E + \lambda_0 \v}
    \left. \int_{C_a} \frac{d \B}{2 \pi i}
      \exp \right[ (\B - \BH) \Eb \no \\
  && \left. + \left( C_{\a} (\B - \BH)^{\a}
    - D_{\a} \T2 (\B - \BH)^{\a +1} \right)
      \v \log \left( \frac{\B - \BH}{\BH} \right) \right],
\EA
and a similar calculation leads us to the same result, that is,
\BE
  \Omega (T,E) \simeq - \frac{4 (2 \pi)^{\a} \v}{{\BH}^{7- \a} \Eb^{\a +1}}
    e^{\BH E + \lambda_0 \v}
      \left( 1- \frac{8 \pi^2 \T2}{\BH \Eb} \right).
\EE
We can calculate the finite temperature effective potential from this, and its $\T2$ term is given by
\BE
  \left[ -16 \tau_p \v
   + \frac{16 \pi^2}{{\BH}^{2} \Eb} \right] \T2.
\EE
From this we can see that the coefficient is always negative, so that a phase transition does not occur.

\end{description}

Let us summarize the results of this section. In {\D{9}} case, the sign of coefficient of the $\T2$ term of the finite temperature effective potential changes from negative to positive near the Hagedorn temperature as the temperature increases, while it remains negative in the case of {\D{p}} with $p \leq 8$. These results lead us to the conclusion that a phase transition takes place only in the case of {\D{9}} pair. Therefore, not the lower dimensional brane-antibrane pair but the space-filling {\D{9}} pair is created near the Hagedorn temperature.

\section{Speculation about the Hagedorn Transition}
\label{sec:Sdual}

Let us turn to discuss the Hagedorn transition in closed string theory. If we consider the closed strings near the Hagedorn temperature, the strings are highly excited and the energy density is very large. Therefore, the effective coupling of strings becomes very large. As a consequence, we must consider this transition in the strong coupling region.

In addition, it seems correct that the coupling of strings becomes very large during this phase transition for the following reason. Let us recall the relation between type II theory and type 0 theory \cite{type0}. The one-loop free energy of type II theory in the high temperature limit coincides with the one-loop amplitude function of type 0 theory. On the other hand, Bergman and Gaberdiel conjectured that type 0A theory is obtained by the orbifold compactification of M-theory \cite{type0dual}. The coupling constant of type 0A theory becomes large as the radius of the compactified eleventh dimension increases. While type 0A theory includes a tachyon field and is unstable, M-theory has no tachyon field and is stable. The above conjecture implies that type 0A tachyon becomes massive at sufficiently strong coupling, and that the tachyonic instability disappears in the strong coupling regime. Therefore, the weak coupling type 0A theory undergoes a phase transition by tachyon condensation and the eleventh dimension decompactifies completely. From this viewpoint one may say that the coupling constant increases unlimitedly during the Hagedorn transition in type IIA theory. For type IIB theory, similar speculation can be deduced from the relation between type 0B theory and M-theory. Thus, we must discuss the phase transition in the strong coupling.

Let us consider the phase transition in type IIB theory. This theory has the S-duality symmetry which interchanges the strong coupling region with the weak coupling one. In the previous section, we have shown that the {\D{9}} system becomes stable near the Hagedorn temperature. If we apply the S-duality to D9-brane, we obtain an NS9B-brane \cite{filling}. The world volume theory of D9-brane is described by open fundamental strings, which live on it. Since the S-dual object of fundamental string is D-string, the world volume theory of NS9B-brane can be described by the open D-strings. D-strings can be treated perturbatively in large $g_s$ region, where $g_s$ is the coupling constant for closed fundamental strings.

Let us see the Hagedorn temperature for D-strings. The Hagedorn temperature for fundamental strings depends only on their tension, which is given by
\BE
  \tau_{\scriptscriptstyle F1} = \frac{1}{2 \pi \ap}.
\EE
On the other hand, the tension of the D-string can be obtained by substituting $p=1$ into (\ref{eq:tension}) as\footnote{Since D-string is a BPS object, we may extend this formula to the strong coupling region.}
\BE
  \tau_{\scriptscriptstyle D1} = \frac{1}{2 \pi \ap g_s}.
\EE
Therefore, we can obtain the Hagedorn temperature for D-strings by replacing $\ap$ in the Hagedorn temperature (\ref{eq:HagedornT}) with $\ap g_s$. The result is
\BE
  {\cal T}_{\scriptscriptstyle DH} = \frac{1}{2 \pi \sqrt{2 \ap g_s}}.
\EE
We can see that the Hagedorn temperature for D-strings is very small when $g_s$ is very large. Our calculation implies that the critical temperature for the phase transition in the case of {\NS9} is slightly below this temperature and it is very small. The reason for this can be understood from the tension of the NS9-brane, which is given by \cite{filling}
\BE
  \tau{\scriptscriptstyle NS9B} = \frac{1}{(2 \pi)^9 {\ap}^5 {g_s}^4}.
\EE
From this we can see that the {\NS9} pair is created in the low energy region when $g_s$ is very large.

From these observations we propose the following model of the Hagedorn transition. Let us start from closed fundamental strings at low temperature. Closed strings are excited as the temperature increases, and the density of strings become very large near the Hagedorn temperature. This lead to the increase of $g_s$, and the tension of NS9B-brane becomes small. Then a phase transition occurs, and the {\NS9} system appears.

Let us make some comments on this model. Firstly, it seems reasonable to assume that higher dimensional NS-charged objects are realized as a configuration of infinitely many lower dimensional NS-charged objects in analogy with the D-brane case \cite{DD}. In this sense, one may say that the {\NS9} system consists of fundamental strings in our model.

Secondly, if we consider the Hagedorn transition in flat spacetime, there is no preferred choice of directions. Thus, it is natural to think that closed fundamental strings change to the spacetime-filling brane, since the spacetime-filling branes manifestly preserve all spacetime symmetries, while the lower-dimensional branes break some of the symmetries \cite{Ktheory2}. Our results seem very reasonable in this sense.

Thirdly, the coupling constant of strings is related to the expectation value of dilaton field. If we denote the expectation value as $\phi$ then the coupling constant is represented as
\BE
  g_s = e^{\phi}.
\EE
Since the dilaton field is one of the massless modes of closed strings, the coupling constant is determined from closed string side. As a consequence, open D-string are kept weakly coupled as long as the closed fundamental string coupling $g_s$ is very large.

\section{Conclusion and Discussion}
\label{sec:conclusion}

In this paper, we have discussed the behavior of the finite temperature effective potential on the brane-antibrane pair in the constant tachyon background. We evaluated the potential at low temperature by using the canonical ensemble method and concluded that the potential minimum shifts towards $T=0$ as the temperature increases in the low temperature region. Near the Hagedorn temperature, we have calculated the $\T2$ term of the potential by using the microcanonical ensemble method. For the {\D{9}} system, the sign of the coefficient of this term changes from negative to positive at slightly below the Hagedorn temperature. This implies that a phase transition occurs at this temperature and the {\D{9}} system becomes stable above this temperature. This result is in sharp contrast to lower dimensional brane-antibrane cases. For the {\D{p}} system with $p \leq 8$, the coefficient remains negative near the Hagedorn temperature, so that such a phase transition does not occur. We thus concluded that not a lower dimensional brane-antibrane pair but a {\D{9}} pair is created near the Hagedorn temperature.

This difference comes from the contribution from the momentum modes of open strings, which can be taken in $p$-dimensional spatial directions on Dp-brane. If we compactify the background spacetime, we must take into consideration not only momentum modes but also winding modes. It would be interesting to investigate the finite temperature effective potential in a compact background spacetime because the energy spectrum of momentum and winding modes depend on the background spacetime. In particular, we might be able to analyze the finite temperature effective action explicitly in the case of torus background, which is the simplest non-trivial one.

Our calculation is based on the one-loop amplitude of open strings, which has been proposed by Andreev and Oft \cite{1loopAO}. As we have mentioned in \S \ref{sec:free}, we have a problem in choosing the Weyl factors in two boundary terms of one-loop world-sheet. Although the choice of Andreev and Oft has the meaning that both side of cylinder world-sheet is treated on an equal footing, we must explain the fundamental reason why this choice is successful even if it is correct. However, we only need to analyze the vicinity of $T=0$ in order to investigate whether the phase transition occurs or not. Our calculation is valid if the mass square is shifted as equations (\ref{eq:massNS}) and (\ref{eq:massR}) for small $|T|$.

We have proposed that the Hagedorn transition in type IIB string theory might be a transition from closed strings to the {\NS9} system. This proposal comes from our calculation and the S-duality. Since the tension of NS9B-brane decreases as the string coupling constant increases, it is natural to think that not only one pair but also large number of brane-antibrane pairs appear in this phase transition. Thus we must consider the finite temperature system of $n$ NS9B-branes and $n$ $\overline{\textrm{NS9B}}$-branes, or their S-dual objects, namely, those of $n$ D9-branes and $n$ $\overline{\textrm{D9}}$-branes.

Type IIA string theory is obtained by applying a T-duality to type IIB theory. Since the T-dual object of NS9B-brane is NS9A-brane \cite{filling}, our proposal is replaced by the phase transition from closed strings to the {NS9A-$\overline{\textrm{NS9A}}$} system in type IIA theory. In strong coupling region, type IIA theory is lifted to eleven dimensional M-theory, and NS9A-brane to M9-brane. Thus, we might have to study the properties of M9-brane in order to investigate the Hagedorn transition in type IIA theory.

If there exist charged objects such as D-branes and NS-branes before the Hagedorn transition occurs, then it seems natural that NS9-$\overline{\textrm{NS9}}$ system includes these objects. Topological configuration of these objects is described by (the S-dual of) the K-theory \cite{Ktheory1} \cite{Ktheory2}. It would be interesting to investigate our model in terms of the K-theory.

The application of our model to cosmology is very important because brane-antibrane pairs might be excited if the temperature of the early universe is close to the Hagedorn temperature. The tension of the brane-antibrane system behaves as the cosmological constant, so that the universe experiences inflation. In fact, much work has been done in this direction \cite{inflation1} \cite{inflation2}.

Finally, the phase transition to the 9-brane is reminiscent of the Plank solid model of Schwarzschild black holes \cite{BH}. It might be interesting to study the black holes as the NS9-$\overline{\textrm{NS9}}$ system.

\section*{Acknowledgements}

The author would like to thank colleagues at Kyoto University for useful discussions. He also would like to thank M. Fukuma for discussions and comments on the manuscript. He is grateful to K. Kikkawa and M. Ninomiya for discussions and encouragement. He appreciates the organizers of Summer Institute 2001 at Yamanashi for hospitality and the Yukawa Institute for Theoretical Physics at Kyoto University, where discussions during the YITP workshop YITP-W-99-99 on "the title of the workshop" were useful to complete this work.

\appendix

\section{Finite Temperature Effective Potential in Canonical Ensemble Method}
\label{sec:canon}

As we commented in \S \ref{sec:comT}, the calculation based on the canonical ensemble cannot be trusted if the temperature is closed to the Hagedorn temperature. In this appendix, we show that the canonical ensemble method gives different results from those obtained by the microcanonical ensemble method, by explicitly calculating the potential based only on the canonical ensemble. From (\ref{eq:highTF}) the free energy is given by
\BA
  F (T, \B) &=& - \frac{4 \v}{{\BH}^{p+1}} \int_{\Lambda}^{\infty} dt \ 
    t^{\frac{p-9}{2}}
      \exp \left( \pi \frac{{\BH}^2 - \B^2}{{\BH}^2} t \right) \no \\
  && + \frac{16 \pi \v \T2}{{\BH}^{p+1}} \int_{\Lambda}^{\infty} dt \ 
    t^{\frac{p-11}{2}}
      \exp \left( \pi \frac{{\BH}^2 - \B^2}{{\BH}^2} t \right),
\label{eq:Fcut1}
\EA
where we have introduced the low energy cutoff. This cutoff is required when we made an approximation to derive (\ref{eq:FHag}). Using the incomplete $\Gamma$ function, (\ref{eq:Fcut1}) can be rewritten as
\BA
  F (T, \B) &\simeq&
   - \frac{4 \v}{\pi^{\frac{p-7}{2}} {\BH}^8 (\B^2 - {\BH}^2)^{\frac{p-7}{2}}}
     \Gamma \left( \frac{p-7}{2} ,
       \pi \frac{\B^2 - {\BH}^2}{{\BH}^2} \Lambda \right) \no \\
   && + \frac{16 \pi \v \T2}
     {\pi^{\frac{p-9}{2}} {\BH}^{10} (\B^2 - {\BH}^2)^{\frac{p-9}{2}}}
       \Gamma \left( \frac{p-9}{2} ,
         \pi \frac{\B^2 - {\BH}^2}{{\BH}^2} \Lambda \right).
\label{eq:Fcut2}
\EA
The potential for each $p$ can be derived from this as follows.

\renewcommand{\descriptionlabel}[1]{\large\bfseries{#1}}
\begin{description}

\item[(a)] {\large \bf \ $p=9$}
\vspace{0.5cm}

If we substitute $p=9$ into the first term of (\ref{eq:Fcut2}), the first argument of incomplete $\Gamma$ function becomes one, so that we can set $\Lambda =0$. For the second term of (\ref{eq:Fcut2}) we can use the following formula for the incomplete $\Gamma$ function;
\BE
  \Gamma (0,x) = -\gamma - \log x
    - \sum_{n=1}^{\infty} \frac{(-1)^n x^n}{n \cdot n!},
\EE
where $\gamma$ is the Euler constant. Combining these two terms, we get
\BE
  F (T, \B) \simeq - \frac{4 \v}{\pi {\BH}^8 (\B^2 - {\BH}^2)}
    - \frac{16 \pi \v \T2}{{\BH}^{10}}
      \log \left( \pi \frac{\B^2 - {\BH}^2}{{\BH}^2} \Lambda \right).
\EE
We can obtain the finite temperature effective potential from (\ref{eq:tacF}), and its $\T2$ term is given by
\BE
  \left[ -16 \tau_9 \v
    - \frac{16 \pi \v}{{\BH}^{10}}
      \log \left( \pi \frac{\B^2 - {\BH}^2}{{\BH}^2} \Lambda \right) \right] \T2.
\EE
This has a similar form to (\ref{eq:p9T2E}) which was derived by using the microcanonical ensemble method. The coefficient vanishes at the critical temperature ${\cal T}_c$ given by
\BA
  {\cal T}_c = \B^{-1} &\simeq& \left[ \BH^2
    + \frac{{\BH}^2}{\pi \Lambda}
      \exp \left( - \frac{{\BH}^{10} \tau_9}{\pi} \right) \right]^{- \frac{1}{2}}
        \no \\
  &\simeq& {\BH}^{-1} \left[ 1
    + \frac{1}{2 \pi \Lambda}
      \exp \left( - \frac{{\BH}^{10} \tau_9}{\pi} \right) \right]^{-1}.
\EA
If we set $\Lambda = (2 \pi)^{-1}$, the result is the same as (\ref{eq:Tc}) which is derived by using the microcanonical method.\footnote{In order to keep the argument of the logarithm in (\ref{eq:Wsingp9}) dimensionless, we have divided $(\B - \BH)$ by $\BH$. But this denominator is ambiguous as long as it has the dimension of the inverse of the temperature. For the different denominator, we can obtain the same result by shifting the cutoff $\Lambda$ in the canonical ensemble method.}

\vspace{0.5cm}
\item[(b)] {\large \bf \ $p:$ even}
\vspace{0.5cm}

In this case we may set $\Lambda =0$ and we obtain $\Gamma$ functions. $\Gamma$ function with half integer argument is given by
\BE
  \Gamma \left( -n + \frac{1}{2} \right) = \frac{(-4)^n n!}{(2n)!} \sqrt{\pi}.
\EE
Substituting this into (\ref{eq:Fcut2}), we obtain
\BA
  F (T, \B) &\simeq&
    - \frac{(-1)^{5- \frac{p}{2}} 2^{10-p} \left( 4- \frac{p}{2} \right) !}
      {\pi^{\frac{p}{2} -4} (8-p)!}
        \frac{(\B^2 - {\BH}^2)^{\frac{7-p}{2}} \v}{{\BH}^8} \no \\
  && + \frac{(-1)^{5- \frac{p}{2}} 2^{14-p}
    \left( 5- \frac{p}{2} \right) !}{\pi^{\frac{p}{2} -6} (10-p)!}
      \frac{(\B^2 - {\BH}^2)^{\frac{9-p}{2}} \v}{{\BH}^{10}} \T2.
\EA
We can obtain the finite temperature effective potential from (\ref{eq:tacF}), and its $\T2$ term is given by
\BE
  \left[ -16 \tau_{p} \v
    + \frac{(-1)^{5- \frac{p}{2}} 2^{14-p} \left( 5- \frac{p}{2} \right) !}
      {\pi^{\frac{p}{2} -6} (10-p)!}
        \frac{(\B^2 - {\BH}^2)^{\frac{9-p}{2}} \v}{{\BH}^{10}} \right] \T2.
\EE
The coefficient remains negative near the Hagedorn temperature, namely $\B \simeq \BH$. Although the result agrees with (\ref{eq:T2p8}) and (\ref{eq:T2p6}) for $p=8$ and $p=6$, respectively, up to numerical coefficient, we obtain different results for $p=4$, $2$ , $0$.

\vspace{0.5cm}
\item[(c)] {\large \bf \ $p:$ odd $(p \neq 9)$}
\vspace{0.5cm}

The incomplete $\Gamma$ function whose first argument is a negative integer can be expanded as
\BE
  \Gamma (-n,x) = \frac{1}{n!} e^{-x}
    \sum_{s=1}^{n} (-1)^{s-1} (n-s)! \ x^{-n+s-1}
      + \frac{(-1)^n}{n!} \Gamma (0,x).
\EE
In our case, $\Lambda \rightarrow 0$ corresponds to $x \rightarrow 0$, so that incomplete $\Gamma$ function can be approximated by the last term. Thus, the free energy can be approximated as
\BA
  F (T, \B) &\simeq&
    - \frac{4 (-1)^{\frac{9-p}{2}}}
      {\pi^{\frac{p-7}{2}} \left(\frac{7-p}{2} \right) !}
        \frac{(\B^2 - {\BH}^2)^{\frac{7-p}{2}} \v}{{\BH}^8}
          \log \left( \pi \frac{\B^2 - {\BH}^2}{{\BH}^2} \Lambda \right) \no \\
  && + \frac{16 \pi^{\frac{11-p}{2}} (-1)^{\frac{11-p}{2}}}
    {\left(\frac{9-p}{2} \right) !}
      \frac{({\BH}^2 - \B^2)^{\frac{9-p}{2}} \v}{{\BH}^{10}} \T2
        \log \left( \pi \frac{\B^2 - {\BH}^2}{{\BH}^2} \Lambda \right).
\EA
We can obtain the finite temperature effective potential from (\ref{eq:tacF}), and its $\T2$ term is given by
\BE
  \left[ -16 \tau_{p} \v
    + \frac{16 \pi^{\frac{11-p}{2}} (-1)^{\frac{11-p}{2}}}
      {\left(\frac{9-p}{2} \right) !}
        \frac{(\B^2 - {\BH}^2)^{\frac{9-p}{2}} \v}{{\BH}^{10}}
          \log \left( \pi \frac{\B^2 - {\BH}^2}{{\BH}^2} \Lambda \right)
            \right] \T2.
\EE
The coefficient remains negative near the Hagedorn temperature either. The result agrees with (\ref{eq:T2p7}) for $p=7$ if we set $\Lambda = (2 \pi)^{-1}$. However, we obtain different results for $p=5$, $3$ , $1$.

\end{description}

\end{document}